# MongoDB Injection Query Classification Model using MongoDB Log files as Training Data

Shaunak S. Perni[a*], Minal Naresh Shiroadkar[b], Ramdas Karmalli[c]

[a] *Goa Business School, Goa University, Taleigao, Durgavado, Goa, 403206, India*
[b] *Goa Business School, Goa University, Taleigao, Durgavado, Goa, 403206, India*
[c] *Goa Business School, Goa University, Taleigao, Durgavado, Goa, 403206, India*

## Abstract

NoSQL Injection attacks are a class of cybersecurity attacks where an attacker sends a specifically engineered query to a NoSQL database which then performs an unauthorized operation. To defend against such attacks, rule based systems were initially developed but then were found to be ineffective to innovative injection attacks hence a model based approach was developed.

Most model based detection systems, during testing gave exponentially positive results but were trained only on the query statement sent to the server. However due to the scarcity of data and class imbalances these model based systems were found to be not effective against all attacks in the real world. This paper explores classifying NoSQL injection attacks sent to a MongoDB server based on Log Data, and other extracted features excluding raw query statements

The log data was collected from a simulated attack on an empty MongoDB server which was then processed and explored. A discriminant analysis was carried out to determine statistically significant features to discriminate between injection and benign queries resulting in a dataset of significant features.

Several Machine learning based classification models using an AutoML library, "FLAML", as well as 6 manually programmed models were trained on this dataset, which were then trained on 50 randomized samples of data, cross validated and evaluated. The study found that the best model was the "FLAML" library's "XGBoost limited depth" model with an accuracy of 71%

All datasets and python notebooks are saved on the following git repository:
https://github.com/ShaunakPerniUniGoa/NoSQLInjectionDetection

---

* Corresponding Author
  Email Address: msci.2105@unigoa.ac.in (Shaunak Perni)
  Phone No: (+91) 8625946258

Generative AI content disclaimer

This paper has utilized generative AI tools, including ChatGPT-4, Grammarly, and Codium, for specific purposes such as code generation, debugging, and linguistic enhancements (e.g., grammar and style corrections). However, the authors affirm that the research, findings, and core content of this paper are solely their own work.

# 1 Background

In 2013, (Son et al., 2013). introduced DIGLOSSIA, a runtime analysis tool for detecting both SQL and MongoDB NoSQL injection attacks in PHP applications. This tool functioned by segmenting database queries into 'code' and 'non-code' components and then analyzing user-supplied input within the 'code' sections. DIGLOSSIA successfully detected all attempted attacks in their benchmark PHP applications. However, its efficacy was contingent on the query strictly adhering to a predefined grammar matching the database's parser, leading to potential false positives or negatives if grammar discrepancies existed. Additionally, its value shadowing mechanism could be incomplete, potentially resulting in false negatives if input-tainted strings were processed by un-monitored third-party extensions or certain built-in PHP functions.

2015 saw two significant contributions. (Joseph and Jevitha, 2015) developed an Automata based NoSQL injection detection system, specifically tested on a Java-based backend with MongoDB. This system achieved a 100% detection rate for 'Time-based' and 'Blind Boolean-based' injection attacks on their testing data, with the authors asserting no false negatives for these scenarios. Nevertheless, its applicability was limited to Java-based systems and these specific attack types, with suggestions for future optimization like using token-level automatons to reduce complexity. In the same year, (Ron et al., 2015) provided a broader examination of security measures for NoSQL databases, confirming that these databases are not immune to injection attacks despite new query and access mechanisms. Their work detailed various attack techniques, including PHP array injections, NoSQL OR JSON injections, and NoSQL JavaScript injections, highlighting how JSON representation, while generally well-defined, permits new types of injection. The study concluded that the NoSQL technological area lagged behind traditional RDBMS SQL systems in comprehensive security measures and awareness. Mitigation strategies recommended included using native encoding tools for JSON queries, DAST/IAST, controlling REST API requests, and diligent implementation of authentication and RBAC.

Moving to 2016, (Hou et al., 2016). further explored the maturity of security measures for MongoDB, analyzing attack and defense at the code level. This research reiterated that NoSQL databases remain vulnerable to injection attacks, with JavaScript or JSON injection principles mirroring traditional SQL injection. Server-side JavaScript injection was experimentally shown to reveal private data. Defense mechanisms discussed included input validation (limiting input to expected formats) and parameterized statements (filtering user input through parameters). The concept of malicious feature detection was also introduced to assess system security. The overall aim was to raise awareness among developers and information security professionals regarding NoSQL security and advocate for security layer implementation.

In 2017 (M. et al., 2017) introduced "NoSQL Racket" a testing tool designed to detect NoSQL injection attacks by comparing the static structure of a NoSQL query in application code with its dynamic runtime statement. This tool leveraged a predefined "Driverstbl" to generate comparable query patterns, with mismatches indicating an injection. "NoSQL Racket" successfully detected attacks across web applications using MongoDB, Cassandra, CouchDB, and Amazon DynamoDB, outperforming commercial scanners in various injection scenarios. Performance testing showed acceptable load handling. However, its reliance on specific integration into application code (as a PHP function) limited its universality as an external scanning tool. The full scope of vulnerabilities addressed, and critical metrics like false positive rates, were not exhaustively detailed, and scalability in very high-throughput environments remained unquantified.

2018 saw two notable contributions. (Eassa et al., 2018). presented DNIARS (Detection of NoSQL Injection Attacks using RESTful Service), an independent RESTful web service for detecting NoSQL injection attacks. DNIARS compared generated patterns from NoSQL statement structures in static and dynamic states to identify injections. Implemented in PHP, its key advantage was the complete separation of the detection process from the web application, enhancing portability and scalability. Evaluation indicated effective operation with a low error rate (not exceeding 1%), and it claimed to detect "all types" of NoSQL injection attacks tested. However, its pattern comparison approach might be susceptible to highly polymorphic or obfuscated attack variations not covered by its internal tables. The broad claim of "all types" lacked comprehensive adversarial testing, and the reported low error rate implied potential for unquantified false positives or negatives. (Sachdeva and Gupta, 2018) focused on basic NoSQL injection attacks on MongoDB, analyzing how these databases, despite different query languages, remain vulnerable due to attackers modifying grammar forms of injection content. They demonstrated basic attacks using PHP and JavaScript on MongoDB and proposed input validation and parameterized statements as defense methods. The concept of malicious feature detection based on a flowchart was also introduced. Acknowledged limitations included the focus on basic attacks, suggesting more advanced possibilities for future examination.

In 2019, (Ul Islam et al., 2019) introduced a tool for automatic detection of NoSQL injections utilizing supervised learning, claiming the development of the first labeled dataset for this purpose. This dataset, comprising 1004 MongoDB queries (203 injections) and 350 CouchDB queries (50 injections), was generated by manually designing features, augmenting queries through cross-overs and mutations, and validating on a vulnerable website. Nineteen features were initially designed, with the top ten selected for effectiveness. Modeling the problem as binary classification, various supervised learning algorithms were evaluated. For MongoDB, the Neural Network classifier achieved the highest mean recall (92.94%) and an F2-score of 0.9343, with the approach demonstrating database-agnostic performance. A comparative study showed their tool outperforming Sqreen by 36.25%. Limitations included the necessity of generating a synthetic dataset due to the lack of publicly available benchmarks, limited literature for dataset enrichment impacting initial size, and challenges in automatic feature design. A larger dataset was suggested for further performance improvement, and a lower variety of "OR injection" examples was noted.

2020 brought an empirical study by (Scherzinger and Sidortschuck, 2020) on the design and evolution of NoSQL database schemas, analyzing ten real-world Java projects from GitHub. This study, claimed to be the largest of its kind, confirmed denormalization as a common practice and observed that NoSQL schemas generally grow in complexity over time, but with higher churn rates (2.8% to over 30% of commits containing schema-relevant changes) compared to relational studies. Additions of entity-classes or schema-relevant attributes were the dominant form of schema change. Limitations included challenges in identifying all schema-relevant attributes from third-party libraries through static analysis, potential under-capture of schema complexity, and inability to programmatically recognize schema changes like renaming or splitting entities. External validity concerns were raised due to the focus on Java projects using specific object-mappers, suggesting a need for broader analysis across languages and NoSQL data stores.

2021 was a prolific year for NoSQL security research. (Shachi et al., 2021) conducted a comprehensive survey on detection and prevention of SQL and NoSQL injection attacks, detailing various NoSQL injection techniques (PHP Tautologies, Union Queries, JavaScript injections, Piggy-backed Queries, Origin Violation) and MongoDB-specific concerns (hash injections, password hashing weaknesses, lack of authentication/encryption in older versions). The survey reviewed both machine learning (ML) and non-ML approaches, noting ML models' reliance on feature-based supervised learning due to the absence of public labeled data. As a survey, it synthesized existing knowledge. Concurrently, Mejia-(Mejia-Cabrera et al., 2021) contributed a new method for constructing a NoSQL query database based on JSON structure for experimental purposes, necessitated by the general unavailability of such databases. They evaluated six classification algorithms, with Neural Network and Multilayer Perceptron achieving 97.6% accuracy. (Alizadehsani, 2021) further advocated for the application of Artificial Neural Networks in NoSQL attack detection, observing that security aspects, particularly injection attacks, have lagged in the NoSQL domain. (Babucea, 2021) provided a review highlighting critical differences and limitations between SQL and NoSQL databases, discussing their evolution, schema flexibility, scalability, and diverse data models. The paper concluded that despite NoSQL growth, no single optimal choice exists, and both models often coexist. Finally, (Weeratunga, 2021) focused on identifying NoSQL Injection Vulnerabilities in MongoDB-based web applications, aiming to automate the process for authentication bypass, unauthorized data access, and JavaScript injection. This research sought to extend the OWASP ZAP DAST tool to support NoSQLi detection, acknowledging the gap in existing DAST tools. The primary limitation was its specific focus on MongoDB and its particular vulnerabilities.

In 2022, (Sajid Ahmed et al., 2022) proposed a two-layer hybrid security firewall for server-side applications to detect both SQL and NoSQL injection attacks in real-time. The first non-ML layer matched incoming request payloads against malicious patterns, blocking immediate threats and reducing the load on the second ML-based layer, which analyzed constructed queries for malicious code. The NoSQL dataset was sourced from earlier research (Ul Islam et al., 2019), and class imbalance was addressed using resampling strategies like SMOTEENN. Feature engineering for NoSQLIA involved extracting 14 features, refined to 12 optimal features. Performance evaluation with algorithms like SVM, AdaBoost, and Random Forest, after SMOTEENN, demonstrated high accuracy (some models achieving 100% accuracy, precision, recall, and F1 score). The system was implemented using Python/Flask and NodeJS/ExpressJS. A primary limitation is its current capability to detect malicious codes exclusively from textual data, with future plans for file type data using deep learning. The 100% reported accuracy might benefit from further validation against highly diverse or adversarial real-world datasets to confirm generalizability and guard against potential overfitting. Praveen et al. proposed a model for NoSQL Injection Detection Using Supervised Text Classification, integrating supervised learning with NLP techniques. This model was designed for multiple NoSQL databases (MongoDB, CouchDB, CassandraDB, Couchbase) and achieved an F1-score of 0.95 through 10-fold cross-validation. However, the abstract did not detail specific injection types or false positive rates, and its robustness against novel or highly obfuscated attack patterns remains an area for more detailed examination.

Most recently, in 2024, (D· l· et al., 2024) introduced "The MongoDB Injection Dataset," a comprehensive collection of MongoDB NoSQL injection attempts and vulnerabilities. Recognizing the increasing vulnerabilities of unstructured databases, they curated 400 NoSQL injection commands (221 malicious, 179 benign) by combining manually authored commands with those acquired via web scraping from sources like Nosqlmap and OWASP. Data augmentation and refinement (filtering commands with Jaccard similarity > 80%) ensured diversity. This JSON-formatted dataset, comprising "text" (payload) and "label" components (malicious/benign), includes complex and simple commands, making it suitable for ML and data analysis, particularly for training or fine-tuning AI models and LLMs to enhance attack detection accuracy. Acknowledged limitations included its tailoring specifically towards MongoDB databases, implying potential limitations in direct applicability or comprehensiveness for other NoSQL database types.

# 2 Introduction

This paper explores the creation of Models to classify log lines of a MongoDB sessions to be either a benign or injection attack. Given the unavailability of a pre-existing labeled dataset of log data for this specific purpose, a synthetic dataset was generated. This involved executing a dataset of both injection and benign queries against an empty MongoDB server and subsequently extracting the corresponding log data. The extracted log data underwent an exploratory phase to identify and retain only significant variables, resulting in a dataset composed solely of the label and these discriminant features.

Classification models were then trained on this processed data. The modeling approach incorporated two distinct sets of models: six manually programmed machine learning models and a model determined by the FLAML AutoML library. To assess generalizable performance, the six manually programmed models were trained, validated, and tested using the same dataset but with varying random seeds. This approach yielded distinct combinations of training, validation, and testing data, from which the average performance of each model was evaluated. The FLAML AutoML model, possessing its own internal system for model selection and performance determination, provided performance metrics directly, which were then compared against the averaged performances of the six manually programmed models.

# 3  Data Collection

(For the code and outputs *Please see ./notebooks/dataProcessing.ipynb on the repository*)

An empty MongoDB 6.0.15 server was set up on a Fedora 39 Linux Machine Kernel version 6.5 with an AMD Ryzen 7 57000U Processor with 8.0 GiB Memory on 1TB storage HDD, using Mongosh 2.25 (a program to interact with the MongoDB database) to send instructions to the database. The profiling level on the server was set to 2 as it was the highest level allowing for the maximum amount data to be recorded in the logs

Using the dataset (D· l· et al., 2024), referred to as the "Query Dataset", in this paper, each query from the dataset was sent to the MongoDB database and then executed. Then, after executing all queries, the log file was extracted. Lines from the log file corresponding to the queries sent were persevered and the rest remaining log line were removed as they had no relevance to the data required for this study. The remaining data was converted from the unstructured JSON format to the tabular form, which is called as "log data" in this paper.

Each nested key, due to the nature of JSON data was then further expanded until each key was an individual feature. The target attribute "label" from the Query Dataset was joined to the log data using the variable "Text" from the Query Dataset and the column "filter" from the log data.

Based on the variable "Filter" from the log data, which contained the raw query string, more features were extracted based on

- Category of operator present in the filter as per MongoDB
- Type of selector present in the filter as per MongoDB
- Presence of a null operand
- Length of the query
- The query with only the MongoDB keywords present and the database variable names removed and the length of the same

After adding these engineered variables all constant variables were removed

The final structure of the collected data is shown in the following table (Table 1).

| Field | Display Name | Data Type | Description |
|---|---|---|---|
| t | Timestamp | Timestamp | Timestamp of the log message in ISO-8601 format. |
| planSummary | Plan Summary | String | Plan used to execute the query. |
| planningTimeMicros | Planning Time in | Float | Time taken to develop a |

|  | Microseconds |  | query plan in microseconds. |
|---|---|---|---|
| cpuNanos | CPU Nanoseconds | Integer | CPU processing time in nanoseconds. |
| filter | Filter | String | Filter used in the query. |
| $eq | $EQ | Boolean | Whether the $eq operator is present in the query filter. |
| $gt | $GT | Boolean | Whether the $gt operator is present in the query filter. |
| $in | $IN | Boolean | Whether the $in operator is present in the query filter. |
| $ne | $NE | Boolean | Whether the $ne operator is present in the query filter. |
| $nin | $NIN | Boolean | Whether the $nin operator is present in the query filter. |
| $type | $TYPE | Boolean | Whether the $type operator is present in the query filter. |
| $mod | $MOD | Boolean | Whether the $mod operator is present in the query filter. |
| $regex | $REGEX | Boolean | Whether the $regex operator is present in the query filter. |
| $where | $WHERE | Boolean | Whether the $where operator is present in the query filter. |
| $elemMatch | $ELEM_MATCH | Boolean | Whether the $elemMatch operator is present in the query filter. |
| $size | $SIZE | Boolean | Whether the $size operator is present in the query filter. |
| $ | Positional Operator | Boolean | Whether the positional $ operator is used. |
| >= | Greater Than or Equal To | Boolean | Whether the query uses a >= comparison. |
| <= | Less Than or Equal To | Boolean | Whether the query uses a <= comparison. |
| < | Less Than | Boolean | Whether the query uses a < comparison. |
| > | Greater Than | Boolean | Whether the query uses a > comparison. |
| selector_comparision | Selector Comparison | Boolean | Whether comparison selectors |

| | | | are used in the query. |
|---|---|---|---|
| selector_logical | Selector Logical | Boolean | Whether logical selectors are used in the query. |
| selector_element | Selector Element | Boolean | Whether element selectors are used in the query. |
| selector_evalutaion | Selector Evaluation | Boolean | Whether evaluation selectors are used in the query. |
| selector_array | Selector Array | Boolean | Whether array selectors are used in the query. |
| selector_bitwise | Selector Bitwise | Boolean | Whether bitwise selectors are used in the query. |
| projection | Projection | Boolean | Whether projection selectors are used in the query. |
| misc | Miscellaneous | Boolean | Whether misc selectors are used in the query. |
| selector | Selector | Boolean | Whether any selector operators are used in the query. |
| standard_logical | Standard Logical | Boolean | Whether standard logical operators are used in the query. |
| all_operators | All Operators | Boolean | Whether all operators are covered in the query analysis. |
| null_operand | Null Operand | Boolean | Whether null operands are used in the query. |
| regex_null_operand | Regex Null Operand | Boolean | Whether regex-based null operands are used. |
| text | Text | String | The filter query as it is |
| query_length_raw | Query Length (Raw) | Integer | Length of the query string as above. |
| keywords_only | Keywords Only | String | The query string but with only MongoDB query keywords |
| query_length_keywords_only | Query Length (Keywords Only) | Integer | Length of the query after extracting keywords only. |
| label | Label | Boolean | Label assigned to the query for identification or categorization. |

*Table 1: Final structure of the collected data*

*Please note due to technical limitations some constant columns remained after automated processing namely "cursorExhausted", "queryFramework" and "reslen" and were removed on manual evaluation*

# 4   Data Exploration

After data collection and processing the dataset. All text based variables were removed and only numerical and boolean/hidden variables were kept as this was the scope of the project. A discrimination analysis was conducted to determine significant variables to discriminate between injection and benign queries.

## 4.1   Discriminant Analysis of Numerical Features

The numerical features "planningTimeMicros" , "cpuNanos" , query_length_raw" and "query_length_keywords_only" were divided into 2 samples based on the label variable i.e. 2 distributions of each variable for injection queries and benign queries. In order to determine if there was a significant difference in the distribution of the features, which meant that there is a difference in the features for an injection and benign query. The distributions of each feature was visualized using a Histogram plot (Figure 1) and a Kernel Density Estimation plot (Figure 2). To confirm if the samples were independent of each other, the Mann-Whitney U test was used with the null hypothesis that the sample means for the feature would not be independent of each other,

The Mann-Whitney test was chosen due to the following reasons:

1. There was no information to suggest that the sample would belong to the same population
2. There was no information to suggest that the samples are dependent on each other
3. There was no information to suggest that the samples are normally distributed nor that the populations they represent would be normally distributed

Since Mann-Whitney U Test does not assume any of these reason it was decided to use this test to determine if the samples came from the same population i.e. the values of the particular feature does not differ for injection and benign queries or if the samples are independent and hence are from different population i.e. the particular feature differs for injection and benign queries

The Results of the Mann-Whitney Test for each features are shown in (Table 2)

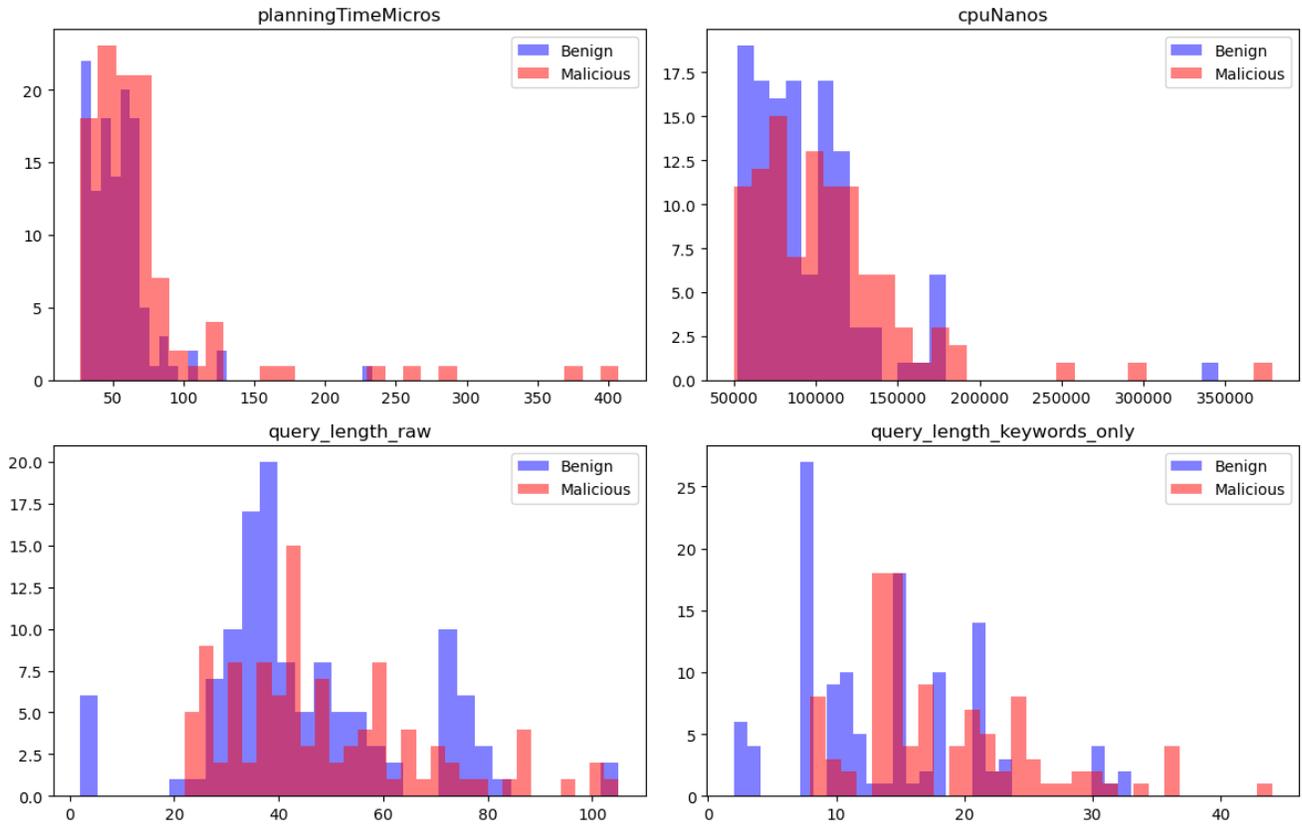

*Figure 1: Histograms of numerical variables, blue represents Benign queries sample and red represents Injection queries sample, clockwise from Top left corner, planningTimeMicros, cpuNanos, query_length_raw, query_length_keywords_only*

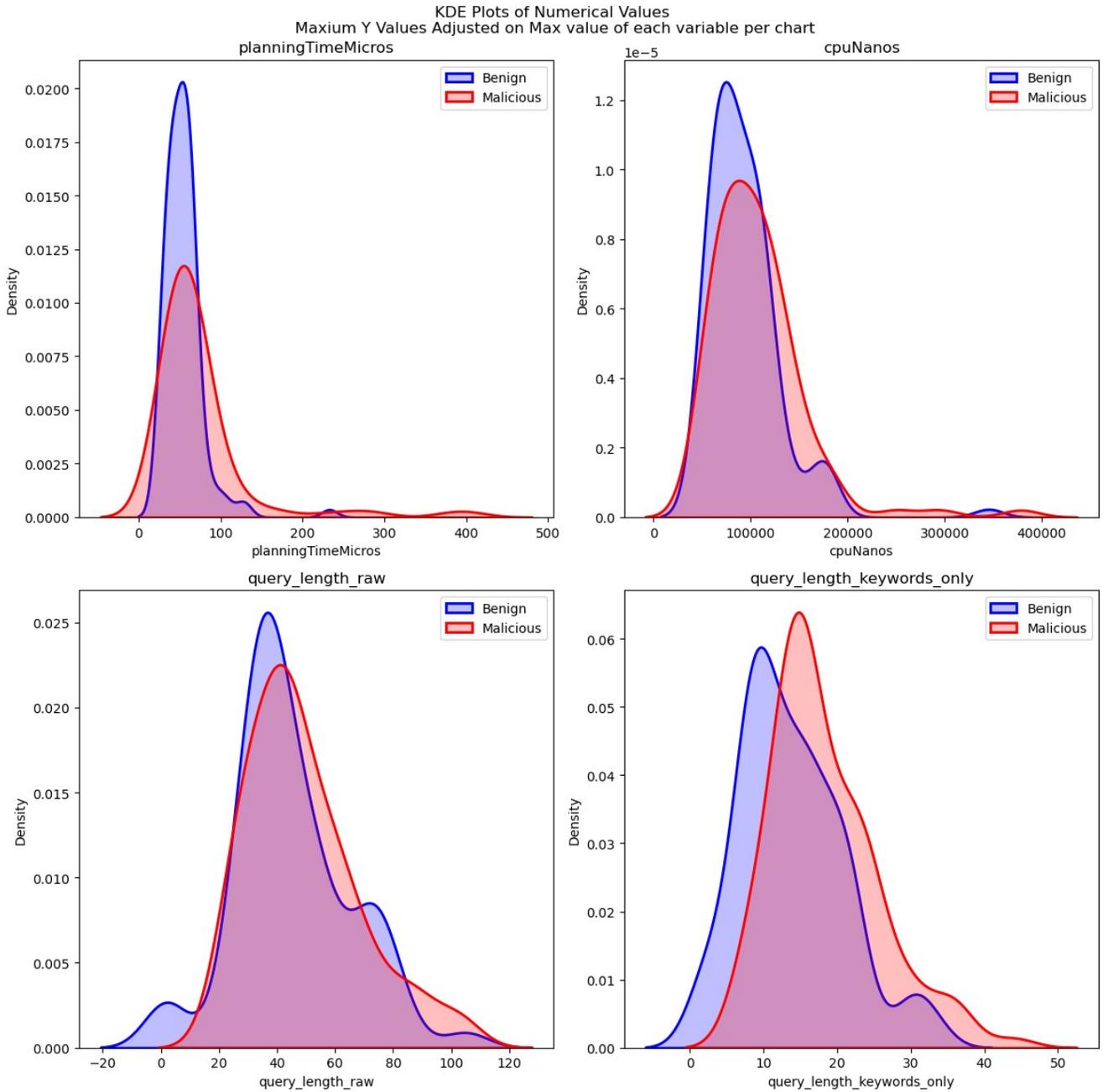

*Figure 2: Kernel density estimation plots of numerical variables, blue represents Benign queries sample and red represents Injection queries sample, clockwise from Top left corner, planningTimeMicros, cpuNanos, query_length_raw, query_length_keywords_only*

| Variable | U Statistic | P-Value | Alternate Hypothesis Accepted at 0.01 Significance |
|---|---|---|---|
| planningTimeMicros | 4926.5 | 0.006628 | True |
| cpuNanos | 5013.0 | 0.011288 | False |
| query_length_raw | 5589.0 | 0.178336 | False |
| query_length_keywords_only | 3397.0 | 0.000003 | True |

*Table 2: Mann-Whitney U Test Results and Statistical Significance at 0.01 Significance for Numerical Variables*

As observed in table 2, only planningTimeMicros and query_length_keywords_only were found to be significant at 0.01 significance hence the samples of these 2 features were confirmed to be significantly different for an injection and benign query

## 4.2  Discriminant Analysis of Boolean Features

For the boolean features, a Chi-Square ($\chi^2$) test of independence was conducted to determine if there was a statistically significant association between each boolean variable and the query type (i.e., whether it was an injection or a benign query). This test was chosen due to its suitability for analyzing the relationship between two categorical variables, without requiring assumptions about the normality of data distribution or the specific shape of the population. The null hypothesis for each test was that the boolean feature was independent of the query type, implying no significant difference in the distribution of the feature between injection and benign queries. The results of the Chi-Square tests for each boolean feature are presented in Table 3.

| Variable | Phi Value | P Value | Alternate Hypothesis Accepted at 0.01 Significance |
|---|---|---|---|
| $eq | 0.08613523235361699 | 0.1973445139252078 | False |
| $gt | 0.055968356591591495 | 0.4022229995323431 | False |
| $in | 0.07056089519199879 | 0.2909413614822473 | False |
| $ne | 0.1727185665577534 | 0.009737481705364296 | True |
| $nin | 0.004795459889571529 | 0.9427834582253255 | False |
| $type | 0.09504298325597839 | 0.15488894901564945 | False |
| $mod | 0.004795459889571529 | 0.9427834582253255 | False |
| $regex | 0.08006523405673249 | 0.23079764009860607 | False |
| $where | 0.11104504640461153 | 0.09651877524572479 | False |
| $elemMatch | 0.11104504640461153 | 0.09651877524572479 | False |
| $size | 0.0369145423668659 | 0.5806152315243337 | False |
| $ | 0.32600281665209946 | 1.0653678148033842e-06 | True |
| >= | 0.004795459889571529 | 0.9427834582253255 | False |
| <= | 0.05437649268574571 | 0.4157407293669543 | False |
| < | 0.05437649268574571 | 0.4157407293669543 | False |
| > | 0.05437649268574571 | 0.4157407293669543 | False |
| selector_comparision | 0.12672397767463786 | 0.05787667883941351 | False |
| selector_logical | 0.12672397767463786 | 0.05787667883941351 | False |
| selector_element | 0.12672397767463786 | 0.05787667883941351 | False |
| selector_evalutaion | 0.12672397767463786 | 0.05787667883941351 | False |
| selector_array | 0.12672397767463786 | 0.05787667883941351 | False |
| selector_bitwise | 0.12672397767463786 | 0.05787667883941351 | False |
| projection | 0.12672397767463786 | 0.05787667883941351 | False |
| misc | 0.12672397767463786 | 0.05787667883941351 | False |
| selector | 0.12672397767463786 | 0.05787667883941351 | False |
| standard_logical | 0.12672397767463786 | 0.05787667883941351 | False |
| all_operators | 0.12672397767463786 | 0.05787667883941351 | False |
| null_operand | 0.12672397767463786 | 0.05787667883941351 | False |
| regex_null_operand | 0.12672397767463786 | 0.05787667883941351 | False |

*Table 3: Phi Coefficients and Statistical Significance for Boolean Variables in Relation to Target "Label"*

as observed in Table 3, only the variables $ne, '$' were found to be significant at 0.01 significance hence these 2 variables were kept and the rest were removed

## 4.3 Separability Analysis

Following the identification of significant variables through the preceding statistical tests, these discriminant features and the label variable were combined to form a 'final dataset.' This dataset was subsequently utilized for separability analysis, with the objective of discerning whether the 'injection' and 'benign' classes exhibited distinct separation within particular dimensions.

The following separability analysis methods were carried out and visualized

1. Linear Discriminant Analysis (LDA) (Figure 3)
   This analysis determines if the data is linearly separable

2. Principal Component Analysis (PCA) (Figure 4)
   This analysis applies a transformation to LDA results and determines if the data is linearly separable

3. t-distributed Stochastic Neighbor Embedding (t-SNE) (Figure 5)
   Configured with two components (or dimensions), this analysis determined the non-linear separability of the data."

### 4.3.1 LDA

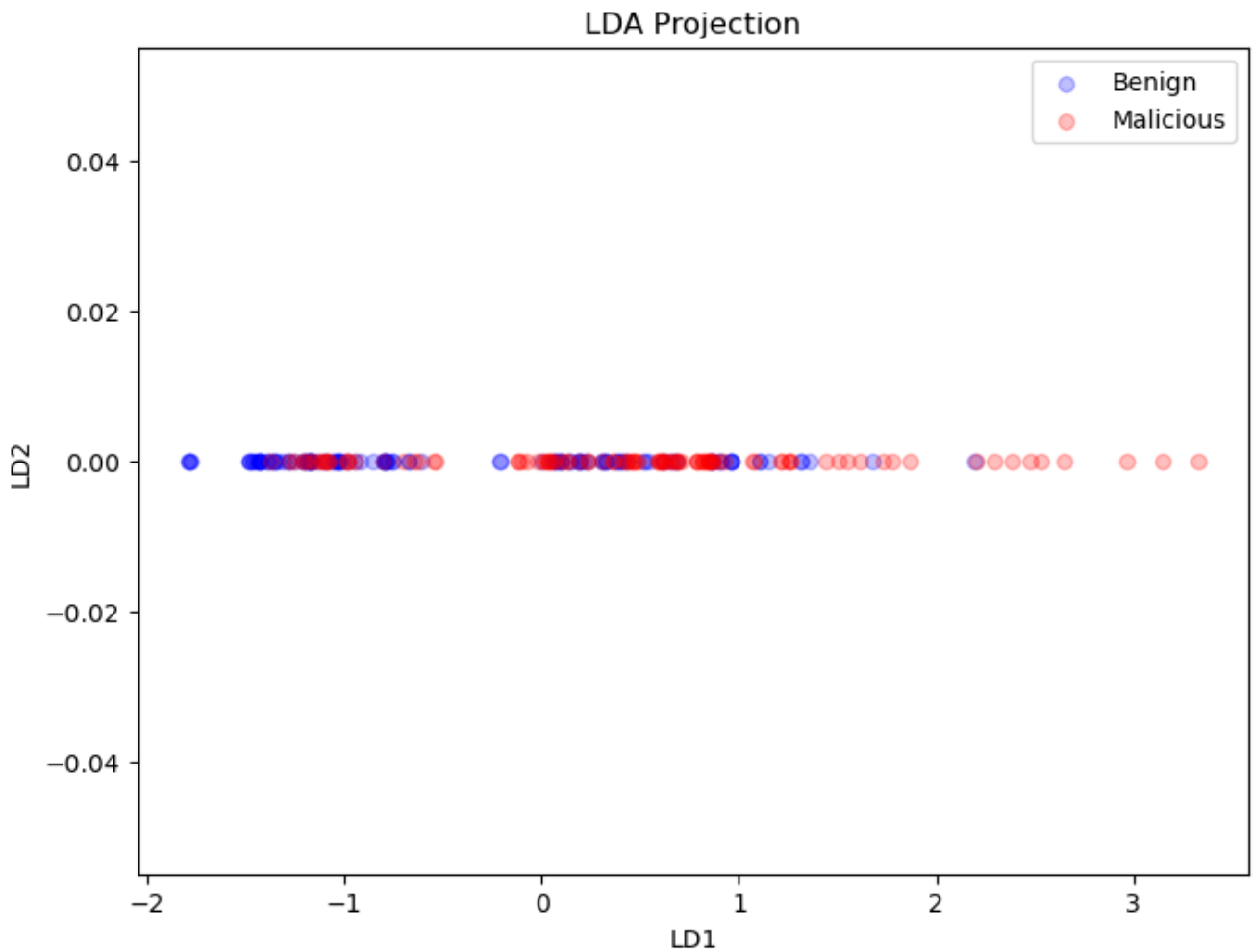

*Figure 3: LDA Projection Graph , Blue circles represent benign query data points and Red circles represent injection query data points*

The following figure shows the LDA visualization

No satisfactory separability was observed in the LDA projection, suggesting that the data may not be linearly separable

### 4.3.2 PCA

The following figure shows the PCA visualization

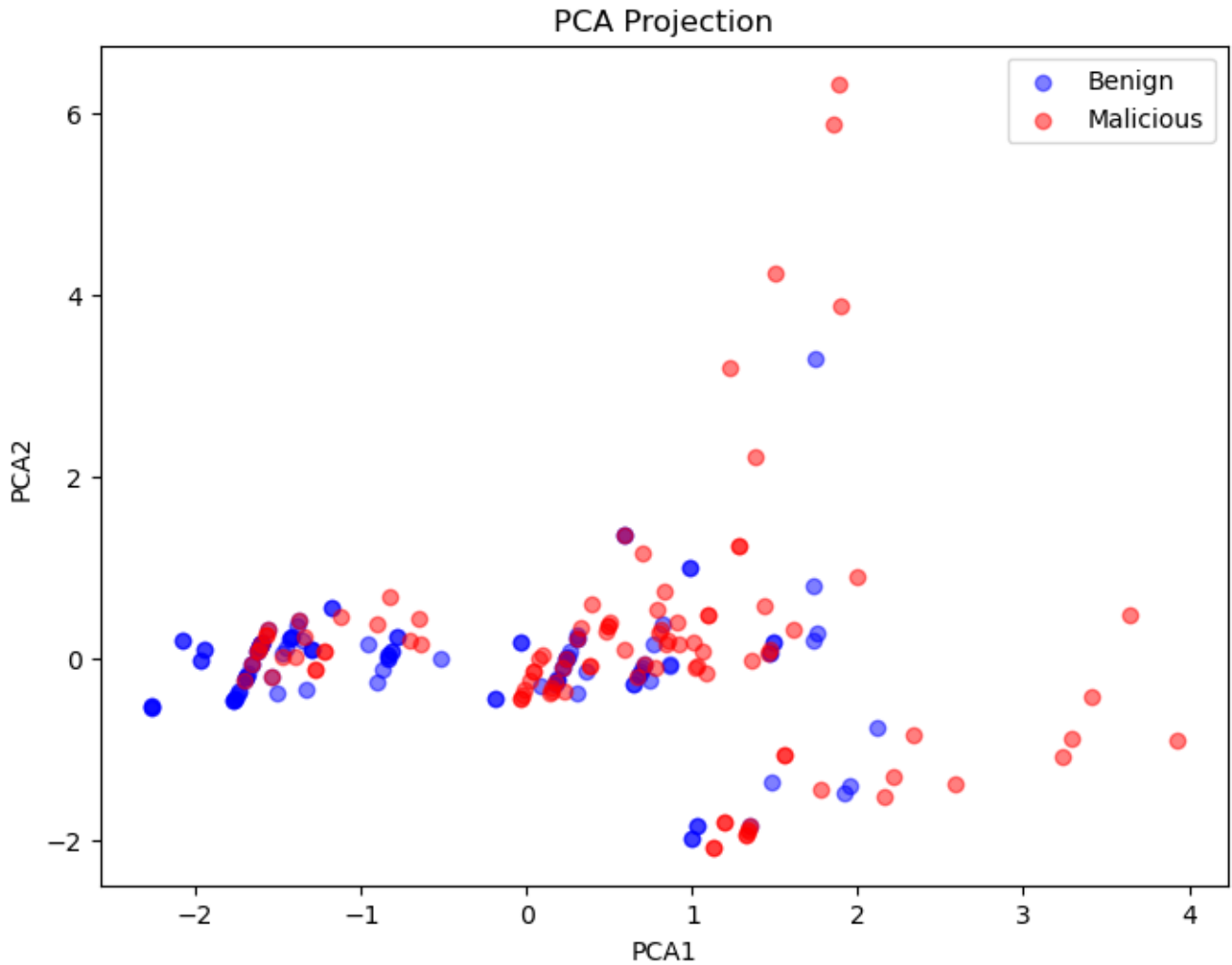

*Figure 4: PCA Projection Graph , Blue circles represent benign query data points and Red circles represent injection query data points*

While the separability in the PCA projection was slightly better than in the LDA projection, it was still not satisfactory, indicating that the data may not be linearly separable.

### 4.3.3 t-SNE

The following figure shows the t-SNE visualization

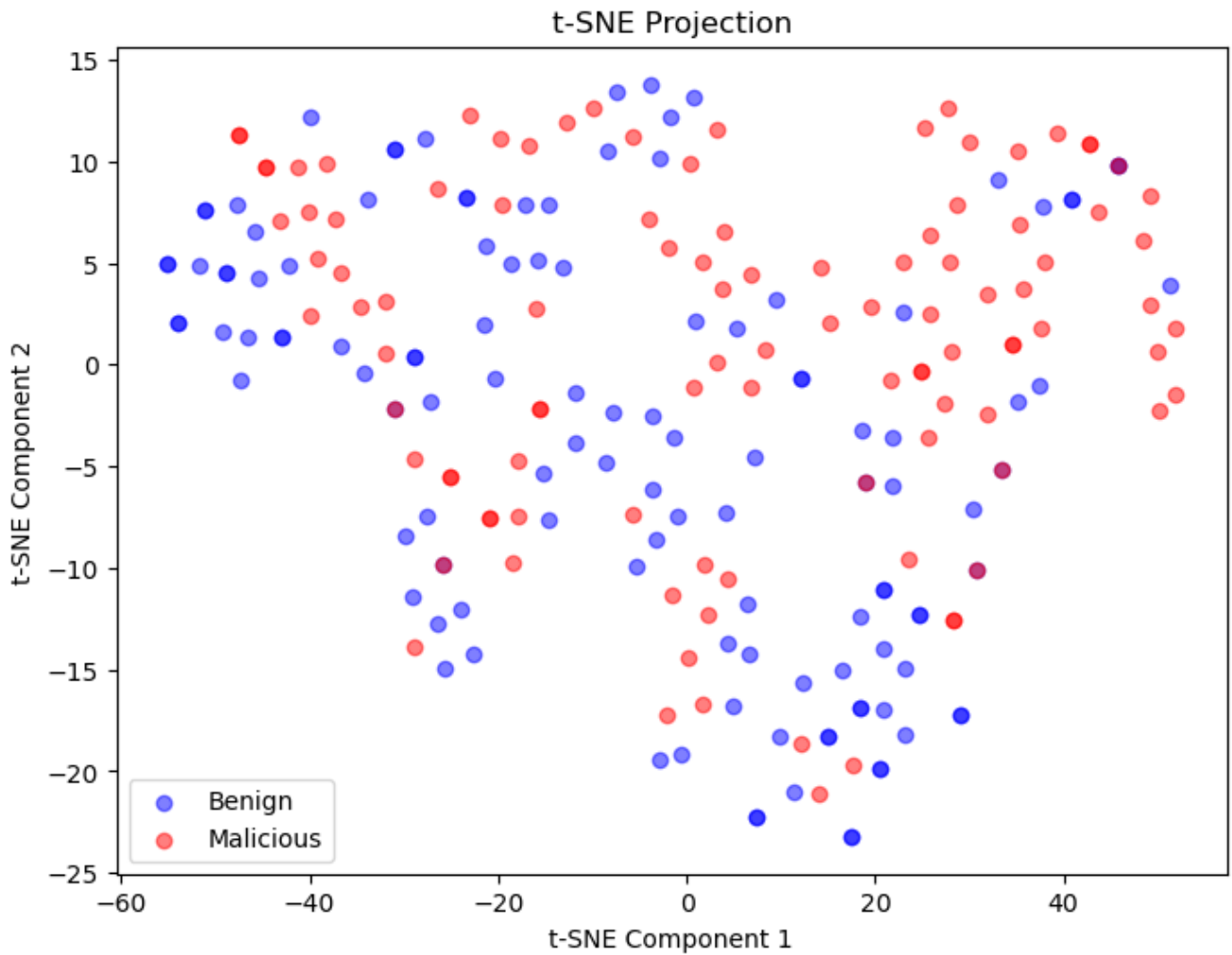

*Figure 5: t-SNE Projection Graph , Blue circles represent benign query data points and Red circles represent injection query data points*

The t-SNE projection showed favorable separability, indicating that the data is separable in a non-linear manner, at least in the 2D reduced space. Thus concluding that linear classification models may not be able to perform well when trained and tested on this dataset and non-linear classification models may perform better

The final dataset's structure with only the target classes and significant features is as follows

| Field | Display Name | Data Type | Description |
|---|---|---|---|
| $ne | $NE | boolean | If a $ne operator exists in the query filter |
| planningTimeMicros | Planning Time Microseconds | float | Time taken to develop a query plan in microseconds. |
| $ | $ | boolean | If the $ character exists in the query filter |
| query_length_keywords_only | Query Length with only Keywords | int | The length of the query filter string if only variables names are removed |
| label | Label | boolean | If the query is a injection query or not |

*Table 4: Structure of data only with statistically significant variables and the target class*

# 5  Model Formulation

A classification model was formulated with the general structure:

Label~ $ne + planningTimeMicros + $ + query_length_keywords_only

This model's objective was to classify queries as either injection or benign based on the specified input properties.

For model evaluation, a set of six distinct machine learning algorithms was selected for training and testing against the prepared dataset:

The algorithms chosen were:

1. Logistic Regression
2. Random Forest
3. Kernel Support Vector Machine
4. K-Nearest Neighbors
5. Decision Tree
6. Naive Bayes

Additionally, the FLAML AutoML model was configured for classification tasks with a time budget of 60 seconds and 5-fold cross-validation (default settings for the model).

## 5.1 Model Training and Testing Overview

The final dataset was systematically partitioned into training, testing, and validation sets, comprising 60%, 20%, and 20% of the dataset, respectively. Models 1-6 were trained on these sets. A loop iterated for random seeds ranging from 1 to 51, generating 50 distinct variations of the dataset splits (X_train, Y_train, X_test, Y_test, X_val, Y_val). For each iteration, the following performance metrics were recorded: Accuracy, Precision, Recall, and F1 Score. Each metric was cross-validated with 5 dataset splits. After all iterations, the average of each metric were calculated for each model. The FLAML AutoML model was provided the training dataset and configured for 5-fold cross-validation (the default value), with results directly recorded as per its internal evaluation methods. The overall training algorithm is depicted in (Figure 6).

## 5.2 Model Training Process

Model training was conducted within a Python 3.13 environment, utilizing the pandas, scikit-learn (sk-learn), and FLAML libraries. The 'final dataset,' loaded as a CSV file, was read into a pandas DataFrame object. This DataFrame was then passed to scikit-learn's `train_test_split` method to generate the six distinct variables for each iteration:

- X_train: 60% of the dataset's unique rows, containing only the significant variables for training.
- Y_train: 60% of the dataset's unique rows, containing only the target variable for training.
- X_test: 20% of the dataset's unique rows, containing only the significant variables for testing.
- Y_test: 20% of the dataset's unique rows, containing only the target variable for testing.
- X_val: 20% of the dataset's unique rows, containing only the significant variables for validation.
- Y_val: 20% of the dataset's unique rows, containing only the target variable for validation.

These variables were used to train, validate, and test the six manually implemented models. Dataset splits were provided to the FLAML AutoML model without external randomization, as FLAML incorporates its own internal system for data partitioning. The iterative process for each random seed ensured comprehensive training and evaluation, with results from each session being recorded and subsequently averaged for overall performance assessment.

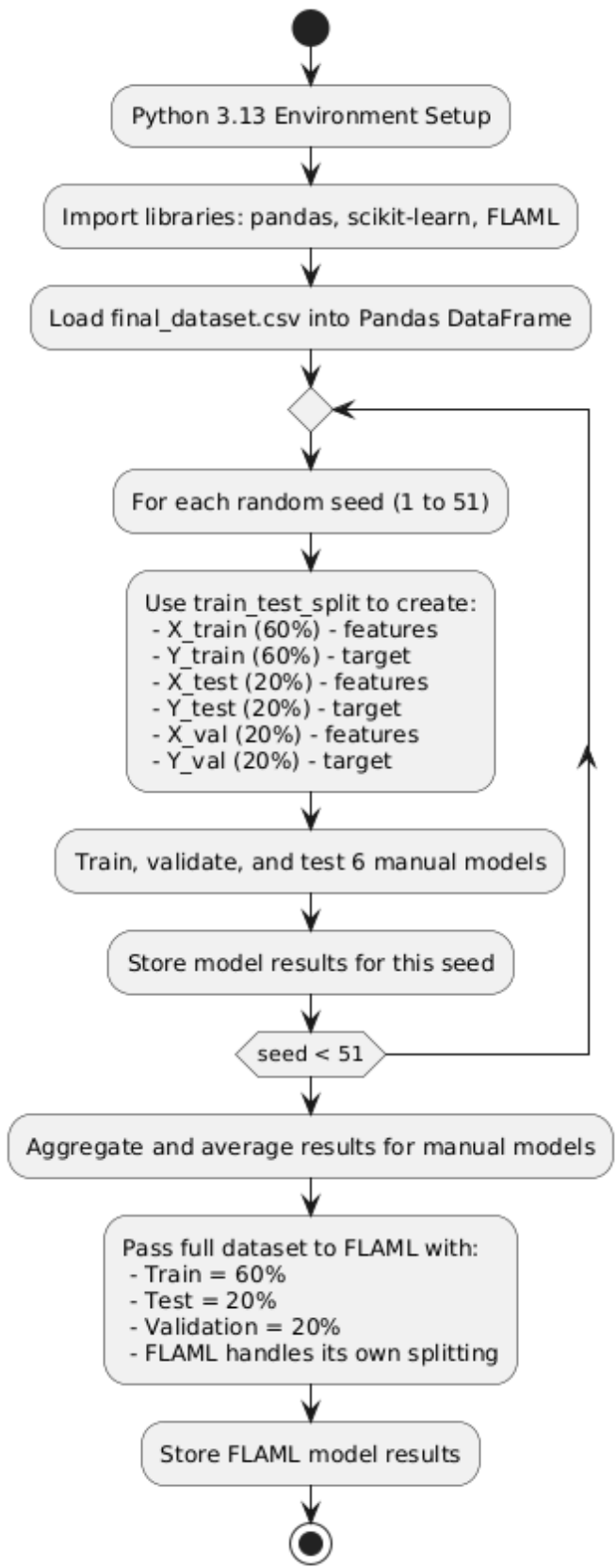

*Figure 6: Model Training Algorithm Flowchart*

# 6   Evaluation

After training and testing and cross validation of the results of all models, the following results were recorded (Table 5)

| Sr No. | Classifier | Accuracy (in %) | Precision (in %) | Recall (in %) | F1 Score (in %) |
|---|---|---|---|---|---|
| 1 | Logistic Regression | 65.45% (avg) | 67.73% (avg) | 66.73% (avg) | 66.64% (avg) |
| 2 | Random Forrest | 67.23% (avg) | 68.38% (avg) | 70.36% (avg) | 67.97% (avg) |
| 3 | Support Vector Machines | 55.97% (avg) | 59.55% (avg) | 46.36% (avg) | 50.00% (avg) |
| 4 | K-Nearest Neighbors | 53.25% (avg) | 53.59% (avg) | 63.27% (avg) | 56.66% (avg) |
| 5 | Decision Tree | 66.32% (avg) | 67.49% (avg) | 66.91% (avg) | 66.71% (avg) |
| 6 | Navie Bayes | 59.87% (avg) | 72.00% (avg) | 39.27% (avg) | 49.28% (avg) |
| 7 | FLAML Best Estimator (XBG Limited Depth) | 73.00% | 75% | 73% | 74% |

*Table 5: Cross Validated and average evaluation metrics of models (1-6) and cross validated and best evaluation metrics of FLAML AutoML model against tested data of different randomization samples*

Hence it was concluded that the XBG Limited Depth algorithm has performed the best with the highest accuracy, precision, recall and F1 score among all other models.

# 7   Conclusion

This project demonstrates the viability of constructing machine learning models to classify MongoDB log entries as either benign or indicative of injection attacks using features derived solely from log data. In the absence of an existing labeled dataset for this task, a fully synthetic dataset was successfully generated by executing both benign and malicious queries against an empty MongoDB instance and capturing the resulting log output. An exploratory analysis phase enabled the identification and retention of only the most discriminative variables, producing a streamlined dataset composed of labels and statistically significant features.

# 8    Limitations

The primary limitation of this study lies in its reliance on synthetically generated data obtained from an empty MongoDB environment. While this approach enabled controlled data collection and precise labeling, it may not fully capture the complexity, noise, and behavioral diversity present in real-world production systems. As a result, the generalizability of the model to live operational environments remains uncertain. Future work should focus on validating these findings using real-world log data and extending the approach to databases containing realistic schemas, workloads, and background activity.